# A Framework for Digital Currencies for Financial Inclusion in Latin America and the Caribbean

Gabriel Bizama, Alexander Wu, Max Mitre, Bernardo Paniagua

**Abstract**
This research aims to provide a framework to assess the contribution of digital currencies to promote financial inclusion, based on a diagnosis of the landscape of financial inclusion and domestic and cross-border payments in Latin America and the Caribbean. It also provides insights from central banks in the region on key aspects regarding a possible implementation of central bank digital currencies. Findings show that although digital currencies development is at an early stage, a well-designed system could reduce the cost of domestic and cross-border payments, improve the settlement of transactions to achieve real-time payments, expand the accessibility of central bank money, incorporate programmable payments and achieve system performance demands.

**Table of Contents**

**Definitions**

<u>Account:</u> A bank or another type of financial institution or mobile money account.

<u>Digital currency:</u> A digital representation of value issued by a central bank or privately-issued which may be transferred and stored electronically, using distributed ledger technology or similar technology.

<u>Digital currency system:</u> A system that involves not only the issuance of a digital currency but also the distribution by financial institutions to consumers and businesses. A digital currency system and a CBDC system are used interchangeably in this report.

<u>Digital payment:</u> A transaction made through a digital channel, including debit and credit card schemes, payments initiated through mobile devices and payments made using digital currencies.

<u>Digital remittance:</u> A remittance sent via a payment instrument in an online or self-assisted manner, and received into an account, i.e. bank account, transaction account maintained at a non-bank deposit taking institution (i.e. a post office), mobile money or e-money account.

<u>Digital wallet:</u> Cash value that is stored on a phone or other electronic device in the form of fiat or digital assets.

<u>Informal economy:</u> All economic activities by workers and economic units that are in law or in practice not covered or insufficiently covered by formal arrangements.[1]

<u>Non-digital remittance:</u> A remittance sent via cash through a physical agent.

<u>Remittance:</u> Household income from foreign economies arising mainly from people's temporary or permanent move to those economies.[2]

---

[1] International Labor Organisation (2013). Decent Work and the Informal Economy. Geneve: United Nations.
[2] The International Monetary Fund, Balance of Payments and International Investment Position Manual, 6th edition (BPM6), 2009.

# 1. Introduction

This research provides a diagnosis of the landscape for financial inclusion and domestic and cross-border payments in Latin America and the Caribbean (LAC) (Section 3). It also provides a framework for assessing the contribution of digital currencies and design considerations to promote financial inclusion in LAC (Section 4). The framework is comprised of the following elements:

- a) Cost and affordability
- b) Availability and settlement time
- c) Accessibility and convenience
- d) Scalability
- e) Programmability

In Section 5, high-level recommendations are provided for the public sector and financial services providers based on the findings. The research employs mixed methods to carefully examine the challenges and opportunities associated with domestic and cross-border digital payments usage and the value of digital currencies in reaching the underserved or last mile populations.

# 2. Methodology and approach

This report synthesizes insights from a survey conducted between the last semester of 2021 and the first semester of 2022 by the Center for Latin American Monetary Studies (CEMLA) with twelve central banks in LAC. It examines quantitative and qualitative data on the access of ownership accounts, domestic and cross-border digital payments usage and costs and key aspects regarding a possible development and implementation of a digital currency system, including stablecoins and central bank digital currency (CBDC). The aim of the report is to provide a framework to assess the contribution of digital currencies for improving financial inclusion.

CEMLA surveyed central banks in LAC to better understand their views on digital currencies, specifically on CBDCs. The questionnaire consisted of 29 questions which are included in Annex 1 of this report. This data is complemented with publicly available data on financial inclusion and payments collected from The World Bank Global Findex, The World Bank Remittances Prices Worldwide, central bank datasets and reports, and further research reports published by international organizations, central banks and the digital currency industry.

The research has several methodological limitations: data collected from public sources on the usage of digital currencies are not representative of the solutions available in the market. Hence, the findings should not be interpreted as representative of all products facilitating payments using digital currencies. For the purpose of this research and report, digital payments are simply defined as transactions made through a digital channel, including debit and credit card schemes, payments initiated through mobile devices and payments made using digital currencies.

# 3. Diagnosis of the current landscape of financial inclusion and payments in LAC

## a) Financial inclusion and the role of payments

**Access to the formal financial system through an account has increased significantly globally in recent years.** The World Bank Global Findex database shows that global account ownership increased from 51 percent to 76 percent between 2011 and 2021.[3] Access to an account is deemed the first step for individuals and micro, small, medium and enterprises (MSMEs) to transact in the formal financial system and, for instance, make or accept digital payments.

**Figure 1: Percentage of global account ownership**

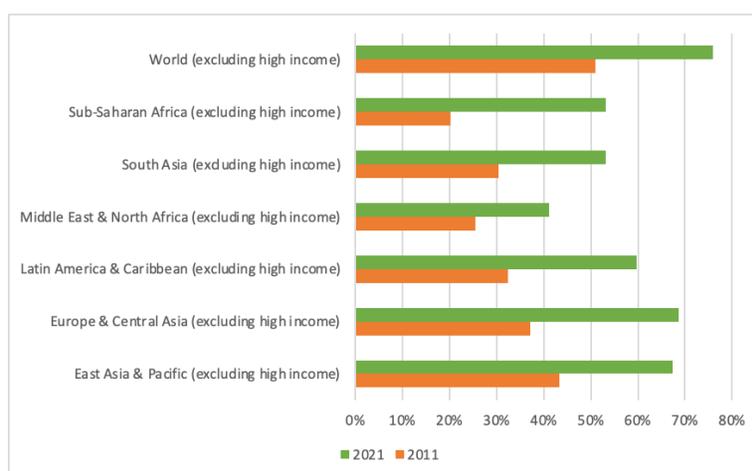

Source: World Bank Global Findex

**Despite increased uptake in global account ownership, LAC countries continue to struggle providing adequate access to accounts and the formal financial system.** Although there has been a significant increase in the ownership of bank accounts globally, data from central banks in the LAC region demonstrates that access to an account is still a challenge. According to survey data collected by CEMLA, while 84 percent of Brazil's population has a bank account, many other countries in the region have substantially lower account ownership rates, which vary between 26 and 38 percent. Financial technology like mobile banking, however, is contributing to improving access to a deposit account. In 2021, access to a mobile money deposit account increased from 5.2 percent in 2017 to 23.4 percent in the LAC region.[4]

---

[3] The World Bank, The Little Data Book on Financial Inclusion, 2022.
[4] Ibid.

**Figure 2: Percentage of individuals aged 15+ with an account in LAC countries**

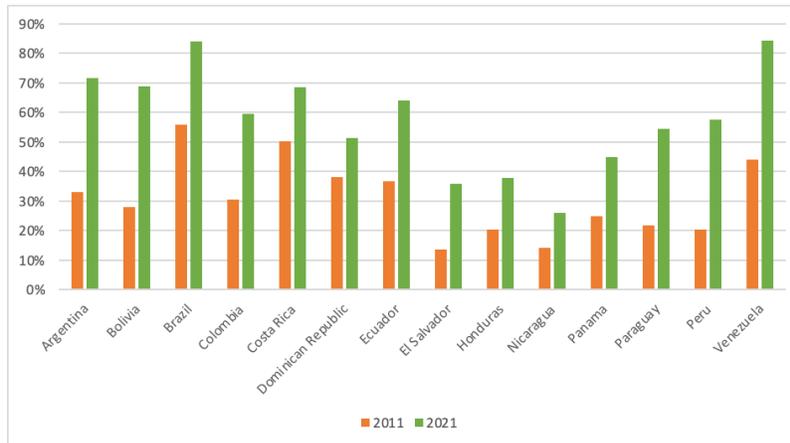

Souce: The World Bank Global Findex

**Figure 3: Percentage of individuals aged 15+ with a mobile money account in LAC**

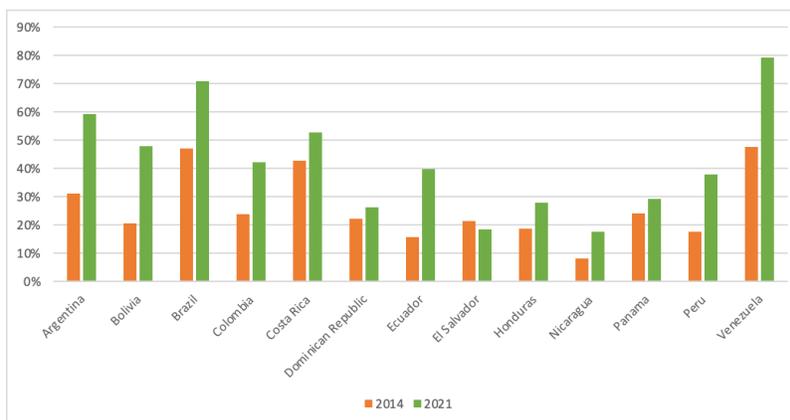

Source: The World Bank Global Findex

**Access to the financial infrastructure and the relevance of interoperability between different payment solutions.** Individuals and MSMEs need to have access to the financial infrastructure in order to use an account. Traditionally, bank branches have been the main gateway to an account but recent digitization of banking services has contributed to improved access. Mobile money through digital wallets has saved cost and time for consumers and has provided easy-to-use solutions for vulnerable groups. However, a challenge that many central banks have worked on is ensuring interoperability among different accounts. Users should be able to transfer funds from a digital wallet with fiat money to a bank account and vice versa and/or make payments using several digital wallets providers supporting different solutions such as QR codes. An open-loop system provides the possibility to make, receive and accept payments using different financial products.

**Digital payments are the entry point to the formal financial system.** Access to the financial system through an account through a bank or mobile money account only is not sufficient for broadening financial inclusion. Access needs to be complemented with usage. A digital payment solution is an important tool for adults and MSMEs to engage with the formal financial system and serve as an entry point to other financial services such as lending, saving, insurance and investments.

**Most countries in LAC have designed, launched and implemented a National Financial Inclusion Strategy (NFIS).** NFISs diagnose and assess the state of financial inclusion and identify priority areas to promote the financial inclusion agenda. NFISs are generally designed and implemented by the relevant authorities which have a mandate to address access to the financial infrastructure, MSME finance, consumer protection, financial education, and inclusive insurance. Digital payments is a core strategy area of NFISs. Most countries in LAC also have a financial inclusion unit within the central bank and/or ministry of finance to lead the agenda and design and implement appropriate policies and regulations to promote the access and usage of formal financial services.

**Figure 5: LAC countries that launched a National Financial Inclusion Strategy**

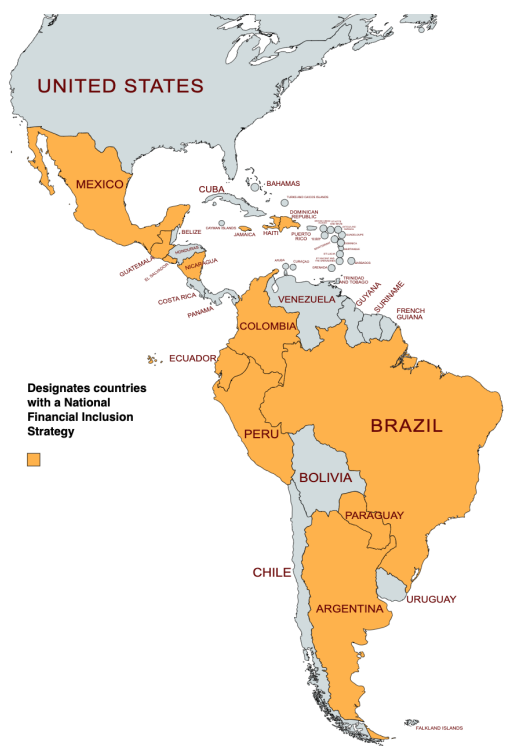

Source: Center for Latin American Monetary Studies and
National Financial Inclusion Strategies Resource Center, The World Bank

### b) Domestic payments

**Usage of digital payments increased significantly in LAC countries, from 45.1 percent to 65.1 percent between 2017 and 2021.** The usage of digital payments both by individuals and MSMEs increased significantly globally and in LAC countries. Data shows that digital payments increased globally from 54.1 percent to 62.1 percent but from 45.1 percent to 65.1 percent in LAC from 2017 to 2021.[5] Central banks are generally in charge of operating the payments systems in each country. For instance, in 2021 the Central Bank of Brazil launched the PIX payment scheme, which connects to the country's instant payment rail (*Sistema de Pagamentos Instantâneos*). There are now more than 3 billion transactions taking place each month, with an

---

[5] Ibid.

average ticket size of about US$88. PIX has also supported financial inclusion by facilitating transactions by 71.5 million individuals (as of December 2022).[6]

**Figure 4: Percentage of individuals aged 15+ that made a digital payment in LAC**

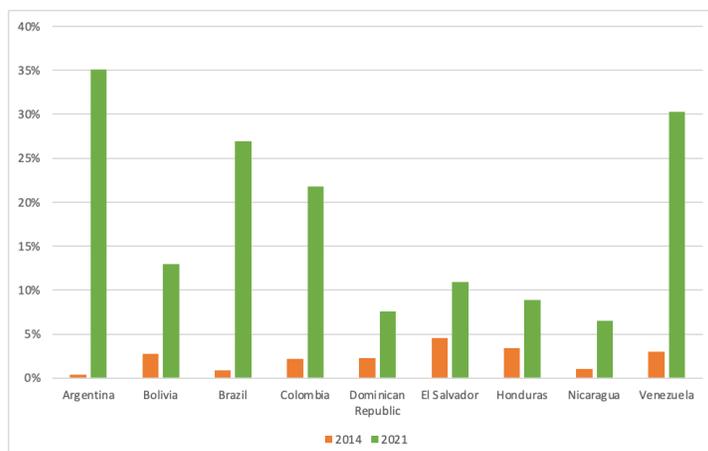

Source: World Bank Global Findex

**Merchant acceptance fees for debit and credit cards remain high.** Although financial regulators and central banks have focused on reducing digital payment fees in recent years, the cost of accepting debit and credit cards remains a barrier to broader usage of digital payments. Data demonstrates that fees in LAC vary from 0.81 percent to 6 percent, depending on the country and whether a debit or card card is used for payments. Significant costs include interchange, merchant acquisition, card issuance, and processing fees. Brazil has the lowest fees at 0.81 percent of total sale for debit payments and 1.92 percent for credit card payments, while Colombia and Argentina have fees as high as 6 percent. Innovative solutions such as mobile money and QR code payments introduce novel digital payment methods and are reducing the overall cost of accepting digital payments.

**Table 1: Average merchant acceptance fee for card payments in Latin America**

| Country | Debit card | Credit card |
|---|---|---|
| Argentina | 3.19% | 5.99% |
| Brazil | 0.81% | 1.92% |
| Colombia[7] | 1.84% | 1.90% |
| Ecuador | 2.24% | 4.50% |

Source: Center for Latin American Monetary Studies

**However, average fees are often not indicative of the true cost that merchants face to accept debit and credit cards.** In Colombia for example, merchants face fees as high as 5 or 6 percent of the cost of goods provided or services rendered. The high fees of the cards scheme reduce the incentive for merchants to digitize their payment methods, which further hinders digital payment adoption.

---

[6] International Monetary Fund, Pix: Brazil Successful Instant Payment System, 31 July 2023
[7] Merchant fees can reach 6 percent, depending on services provided.

**Table 2: Merchant payment acceptance fees by industry in Colombia**

| Industry | Debit card | Credit card |
|---|---|---|
| Acquirer: MCCA | | |
| Automotive | 0.0-5.7% | 0.0-4.5% |
| Pharmaceutical | 0.85-5.1% | 1.2-4.5% |
| Acquirer: Credibanco | | |
| Education | 0.56-5.1% | 0.56-5.1% |
| Books and Print | 0.0-6.1% | 0.89-6.1% |
| Acquirer: Redeban | | |
| Medical Services | 0.55-4.5% | 0.83-6.2% |
| Professional Services | 0.0-5.6% | 0.0-4.4% |

Source: Grupo Bancolombia

**Box 1. Cost of peer-to-peer (P2P) transfers, merchants mobile point-of-sale, and QR payments in Argentina**

In 2022, the volume of payments transactions per adult increased by 54 percent, and the transaction amount by 25 percent in real terms. Adults made an average of 17.8 monthly payments, a new record that is virtually three times as high as the pre-pandemic figure. These increases were driven in part by innovative solutions that lowered the costs for merchants to accept mobile payments.

Cost of accepting a payment initiated with a mobile device by a merchant

The cost for a merchant to accept a P2P transfer initiated with a mobile phone amounts to 0.8 percent and a payment with a debit card associated with a mobile device amounts to 1.3 percent.[8]

Mobile point-of-sale (mPoS)

In 2022, the number of mPoS devices increased 17 percent amounting to 4.4 million devices, a record high in the series. The significant expansion in the number of mPoS devices can be explained by low acquisition and maintenance costs as well as ease of movement and use.

P2P transfers initiated with a mobile device

Two out of three transfers per adult were initiated by a mobile device (from a payment account and bank transfers made through mobile banking). This likely reflects an increased preference to make everyday transactions for relatively small amounts from a mobile phone. This trend is correlated with more extensive use of mobile phones (88.1 percent of the population)

Payments by transfers (PCTs)

PCTs are instant transfers to pay for goods and/or services. In 2022, every adult averaged 3.5 PCTs monthly, which represents a 123 percent increase. These transactions occur among clients of the same PSP with funds available at payment accounts, and are initiated through a QR code or a payment button. A PCT acquirer provides tools for merchants to receive PCTs.

Interoperable PCTs are instant transfers initiated by various financial services providers using interoperable QR codes and card credentials and payment buttons for PoS terminals. Interoperable PCTs per adult increased 64 percent, mostly driven by transactions initiated through QR codes. The average value per transaction fell, which suggests more widespread use for everyday transactions.

*Source:* Banco de la República Argentina, Informe de Inclusión Financiera, April 2023.

**Creating incentives to increase consumer usage and merchant acceptance of digital payments.** Central banks, ministries of finance and other relevant authorities have also designed incentives to encourage increased usage of digital payments by consumers and acceptance of those payments by retailers. Incentives include lottery schemes, PoS terminal subsidies and tax rebates. Evidence shows that the impact of these incentives is high in cash-based societies in the short term because these novel incentives trigger new behavioral responses. However, their impact diminishes in the long term when individuals revert to the preexisting behaviors of their day-to-day routine in the absence of incentives.[9]

---

[8] Interchange rate in Argentina for payments initiated by mobile devices according to local regulations.
[9] The World Bank, Financial Inclusion Global initiative, Incentives for Electronic Payments Acceptance, 2022.

**Box 2. Incentives to increase merchant acceptance payments: Uruguay, Colombia and Mexico**

<u>Uruguay and Colombia: Tax rebates and PoS terminal subsidies</u>

In 2014, Uruguay enacted the Financial Inclusion Law, which provides for value-added tax (VAT) reductions for every retail purchase made with a digital payment. The law sets a timetable for periodic VAT reductions but the Ministry of Finance decided to maintain a rate of 4 percent after it observed in 2017 that the initiative worked beyond expectations. Colombia also implemented VAT reductions for consumers that make digital payments. Even though VAT reduction policies were not complemented with other policy measures such as PoS terminal subsidies, the policy measures had positive results initially. However, it frequently took local tax administrations one to three months to identify deductible transactions and issue refunds via bank accounts.[10] This indicated that VAT rebate systems were likely complex and burdensome for merchants to process.

Subsidies on PoS terminals are another incentive to drive merchant acceptance of digital payments. For example, in Uruguay a tax credit was applied to a business' future tax payments if it installed a PoS terminal. In addition, the government also subsidized the rental cost of terminals for some merchants.

<u>Mexico: Lotteries</u>

A lottery scheme called 'El Boletazo' was implemented in Mexico in 2004 with the intention of reducing cash usage. Adults making digital payments were entitled to participate in lotteries where they could win prizes, including cars and washing machines. Raffles were broadcasted daily on national television and supported by payment services providers Visa and Mastercard. Although the scheme was popular, an impact assessment demonstrates that it did not change payment behaviors in the long term.[11] Similar schemes were introduced in other countries such as Korea, Italy, Kenya and India with similar results.[12]

### c) Cross-border payments

**Cross-border payments are crucial for enabling international commerce and economic development; however, they are still costly, slow and not accessible 24/7.** Both remittances and business-to-business payments are important components of global trade and economic growth. Despite the continued efforts of the international community, cross-border payments remain expensive, slow, and not accessible 24/7. Efficient cross-border payments solutions in terms of affordability and speed can advance financial inclusion.

**Remittances are a critical source of funds for countries in LAC.** This is due to the value of remittances relative to the size of their economies and the high total value of the transfers. Remittances represent a significant source of foreign exchange for most LAC countries, reaching levels comparable to those of foreign direct investment. For example, in El Salvador, Haiti, Honduras, and Jamaica, remittances represent at least 20 percent of their gross domestic product (GDP). In 2022, Mexico remained the largest remittance recipient with $60.3 billion, followed by Guatemala with $18.1 billion, the Dominican Republic with $9.9 billion, and

---

[10] Ibid.
[11] Castellanos, S. G., Garrido D., "Tenencia y uso de tarjetas de crédito en México. Un análisis de los datos de la encuesta nacional de ingresos y gastos de los hogares 2006," El Trimestre Económico, 305, 69-103.
[12] The World Bank, Financial Inclusion Global initiative, Incentives for Electronic Payments Acceptance, 2022.

Colombia with $9.1 billion, while Honduras and El Salvador received $7.3 and $7.6 billion, respectively.[13]

**The cost of sending funds across borders still remains high with a global average cost of 6.25 percent for every USD 200[14].** The global average cost of remittances increased 0.01 percentage points from 2022 Q4 to 2023 Q1. This is more than double the Sustainable Development Goal (SDG) target of 3 percent by 2030.[15] The average cost of sending USD 200 for digital remittances is 4.72 percent. These figures only consider the operational costs of sending remittances incurred by service providers and do not consider additional taxes levied by recipient countries.

**Figure 6: Global average cost of remittances**

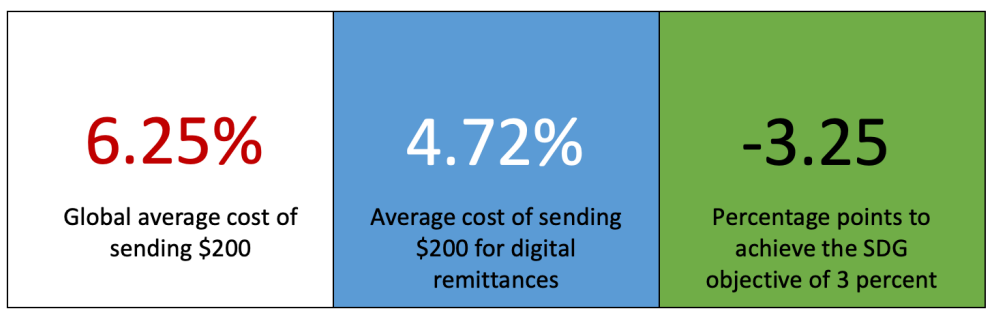

Source: The World Bank, Remittances Prices Worldwide

**Technology contributes to a reduction in the average cost of remittances.** Digital remittances involve sending money electronically. The digital transfer of remittances has proven to be an effective method to significantly reduce the cost of sending money globally due to the reduction of intermediaries such as correspondent banking, liquidity providers, among others. The global average for digital remittances is lower (recorded at 4.72 percent) than non-digital remittances (recorded at 6.92 percent).[16] Data demonstrates that the growing development of digitalization allows for more affordable solutions and products that could advance financial inclusion.[17]

**Figure 7: Trends in the global cost of sending $200 in remittances**

---

[13] United Nations Capital Development Fund, South-South Learnings on Analyzing Remittances Data: Experience Across Central Banks in Latin America and the Caribbean, 2023.
[14] The World Bank, Remittances Prices Worldwide Quarterly, 2023.
[15] United Nations, Sustainable Development Goals, 2023.
[16] The World Bank, Remittances Prices Worldwide Quarterly, 2023.
[17] International Monetary Fund, Digital Financial Inclusion in Emerging and Developing Economies: A New Index, 2021.

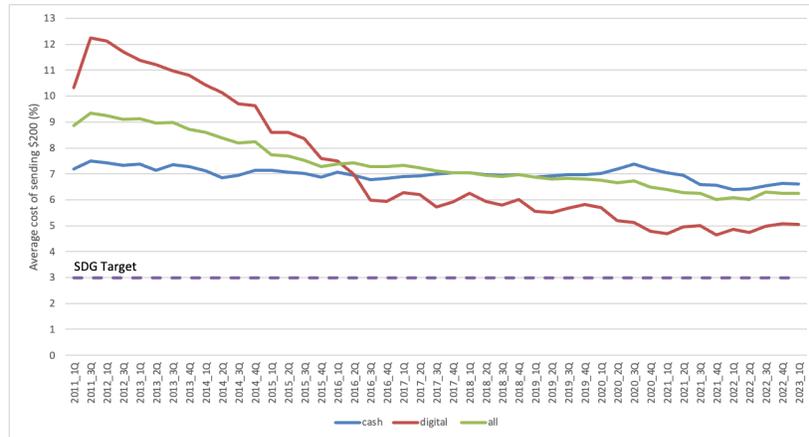

Source: The World Bank, Remittance Prices Worldwide

**The cheapest method for funding a remittance transaction is mobile money; however, it is still higher than the 3 percent SDG target.** In Q1 2023, the cheapest method for funding a remittance transaction was mobile money at 4.36 percent. Sending money using a credit or debit card costed 5.34 percent. The average cost of cash remittances was 6.60 percent. Using a bank account incurred an average cost of 7.45 percent. The cost of sending remittances to a bank account within the same bank or to a partner of the originating bank amounted to 8.07 percent in Q1 2023. In contrast, sending money to a bank account regardless of originating bank was 6.78 percent. When funds are sent to a mobile wallet the average cost was 4.31 percent. Services where money is disbursed in cash costed on average 6.15 percent.[18]

**Figure 10: Average cost by instrument used to fund the transaction**

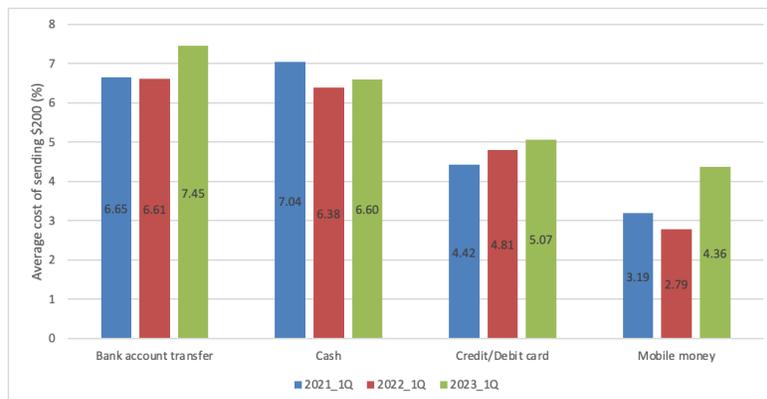

Source: The World Bank, Remittance Prices Worldwide

**Figure 11: Average cost by means of disbursing the funds**

---

[18] The World Bank, Remittances Prices Worldwide Quarterly, 2023.

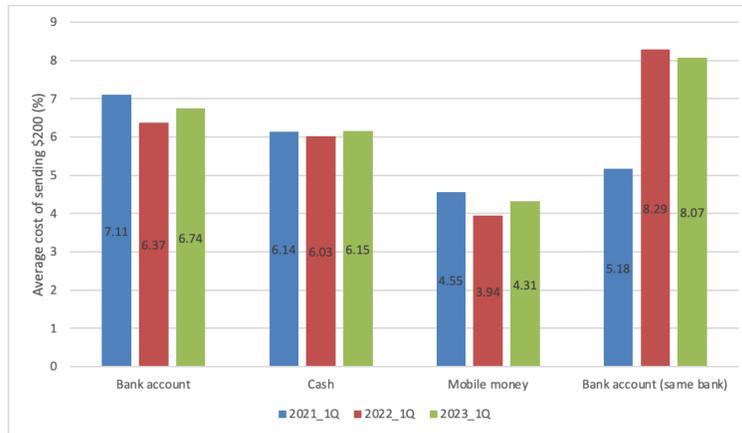

Source: The World Bank, Remittance Prices Worldwide

**Mobile operators are the cheapest remittance service providers.** Competition in the remittances market has led to a broad range of financial services providers offering and facilitating international payments. Although banks and post offices have decreased the cost of cross-border payments over time, they remain firmly above the global average of 6.25 percent. Mobile transfer operators (MTO) and mobile operators have remained below the global average and are contributing to more affordable remittance methods.[19]

**Figure 12: Total average over time by remittance service provider**

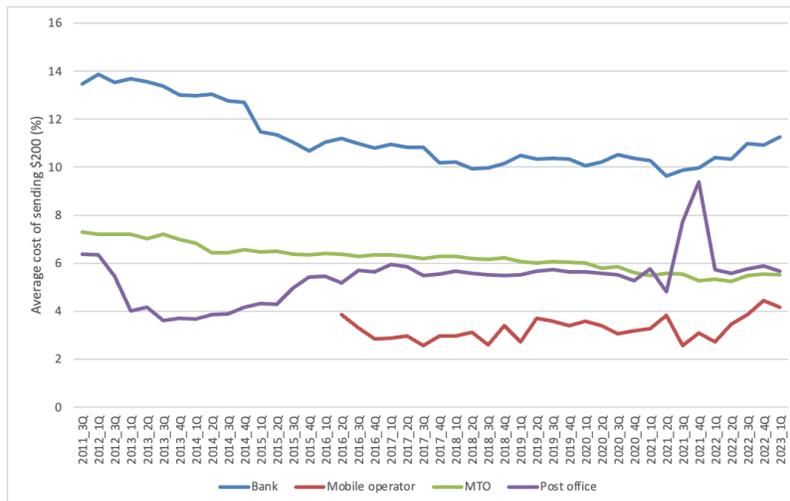

Source: The World Bank, Remittance Prices Worldwide

### d) The case for CBDCs

**Central banks in LAC have expressed interest in developing a retail CBDC and highlighted financial inclusion as a potential policy goal for a retail CBDC.** In the 2022 CEMLA survey, retail CBDC development was identified as an area of interest over wholesale CBDC. Financial inclusion was commonly cited as a motivation for central banks in LAC to explore CBDCs, in part because promoting financial inclusion is a key policy mandate for some central banks, also considering the current state of financial inclusion in LAC. According to the

---

[19] The World Bank, Remittances Prices Worldwide Quarterly, 2023.

World Bank's Global Findex, account ownership in the region amounts to 72.90 percent and digital payments made or received in the last year amounts to 65.10 percent.[20]

**Cash is still used predominantly in the region, which hinders financial inclusion efforts; a CBDC could reduce the use of cash and make payments more affordable both for consumers and merchants.** There is still widespread use of cash in the LAC region, mainly in rural areas. New innovative payment solutions such as QR code payments have reduced the use of cash by adults and MSMEs; however, other barriers like operational costs and tax burdens suppress access to the formal economy and preclude further growth and development of the use and acceptance of digital payments. Other non-pecuniary costs to cash transactions like carrying banknotes and relying on intermediaries to deliver cash also can be addressed with digital payment solutions. Some central banks highlighted that a CBDC could reduce the use of cash and make payments safer, faster, and more affordable.

**CBDCs may contribute to increasing the distribution and access of central bank money in remote areas and reducing the cost of printing central bank money.** Several countries face the challenge of making central bank money available to the population. Access to cash in remote areas is expensive, as is the case for countries in the Caribbean comprising several islands. Common solutions like automated teller machines (ATMs) are difficult to integrate in geographically dispersed regions like the Caribbean island nations, especially when connectivity is inconsistent. Central banks highlight that CBDCs may address this challenge by increasing distribution of central bank money and creating more use cases through features like offline payments. CBDCs can also contribute to reducing the cost of printing central bank money. A central bank in  that the total cost of currency management is estimated at approximately 1 percent of GDP. Central banks do acknowledge that a CBDC requires major infrastructure development, which could be expensive.

**Connectivity is a challenge in remote areas.** For a CBDC to be implemented successfully and achieve a high level of adoption, access to digital infrastructure is needed, particularly in remote and rural areas in LAC. In addition, connectivity should be complemented with financial literacy. Several central banks in the region identified connectivity and financial literacy as the most relevant challenges to address in the potential implementation of a CBDC.

**Central banks are in an early stage of CBDC development; some of them have explored issuing a CBDC using blockchain technology.** Although several central banks in LAC are exploring CBDC implementation and have tested solutions, most central banks in LAC remain in the investigation and research phase with the purpose of understanding opportunities and challenges. Some central banks have created working groups to identify the needs based on the local context. Other central banks have initiated the exploration and development of CBDC solutions using blockchain as the underlying technology.

---

[20] The World Bank, The Little Data Book on Financial Inclusion, 2022.

**Summary of main results**

| Main Results | |
|---|---|
| **Common points** | **Challenges** |
| There are several common topics of interest that can facilitate cooperation among central banks of the region. | Each country has its own characteristics, resources and policy needs to address. Therefore, the design and implementation of a CBDC should be tailored to the specific context of each jurisdiction. |
| One of the most common motivations to implement a CBDC is financial inclusion. | Financial inclusion in the region amounts to 72.9 percent. Financial inclusion is one of the top priorities for some central banks in the region. |
| Reasonable adoption rate of digital payments in certain sectors of the population. A CBDC could trigger further adoption of digital payments. | Culturally, there is a preference for cash and lack of trust in electronic money. |
| The cost of printing and distributing central bank money could be reduced by the implementation of CBDCs. | A CBDC requires a major infrastructure development, which will also be expensive. |
| A CBDC can improve the cost and speed of cross-border payments. | Interoperability among CBDCs in different jurisdictions. |
| Distributing central bank money in remote areas is expensive. CBDCs may address this challenge. | CBDCs require internet connectivity. Communities in remote and/or rural areas often lack access to digital infrastructure since mobile phones often do not possess the technical requirements to support a digital wallet, among other digital financial services.. |
| Central banks are the sole authority for minting and distributing central bank money and to maintain and update the financial system. | The implementation of a CBDC could require an amendment to the current legal and regulatory framework in jurisdictions. |
| From a technical implementation perspective, there is a common expectation that blockchain technology could be beneficial to address policy goals in relation to payments. | Blockchain experimentation is in very early stages. |

## 4.    A framework for digital currencies for expanding financial access and usage.

This section considers a framework for assessing the contribution of digital currencies to promote financial inclusion in LAC. The framework is comprised of the five following elements:

| Element | Parameter | Metric |
|---|---|---|
| a) Cost and affordability | Domestic: Is a digital currency solution contributing to further reduce the cost of merchants acceptance of payments? | Domestic: Average cost of acceptance fees for merchants should be below 0.8 percent. |

|  | International: Is a digital currency solution contributing to further reduce the cost of cross-border payments? | International: Average cost of cross-border payments should be below 2 percent. |
|---|---|---|
| b) Availability | Would a digital currency system expand the hours of operations of a national payments system? | A digital currency system should be available 24/7. |
| c) Accessibility and convenience | Would a digital currency system contribute to expanding the access to money? | A digital currency system should be accessible in remote and rural areas |
| d) Scalability | Would a digital currency system process a higher volume of transactions per second than the current national systems? | A digital currency system should process the volume of transactions per second required by the performance demands defined by each national payment system operator in LAC. |
| e) Programmability | Would a digital currency system enable programmable payments under specific conditions? | A digital currency system should possess the flexibility to tailor the distribution of digital currencies to incorporate specific conditions required by the local context. |

### a) Cost and affordability

**Domestically, digital currencies could contribute to reducing the cost of merchant acceptance of digital payments.** Although there are currently some CBDC implementations globally, there is no clear evidence that suggests that digital currencies are reducing the fees for accepting digital payments by retailers. However, this is an area to further explore in the design and implementation phases of digital currency systems.

**Internationally, digital currencies are contributing to further reduce the cost of cross-border payments.** The total cost of cross-border payments using digital currencies is estimated at 2 percent. The breakdown of fees is divided into two main components: on- and off-ramp fees and exchange fees. On- and off-ramp fees are typically commissions paid to financial entities such as wallets that facilitate the conversion of fiat to digital currencies like stablecoins. These fees vary between 0 to 1 percent. Exchange fees are another variable if there is insufficient liquidity at the time of transaction in a specific payment corridor. Liquidity provider fees amount to 0.01-0.05 percent.[21]

**Table 3: Total cost and breakdown of fees for sending cross-border remittances using stablecoins**

| Fees and costs | Description of fees and costs | Range estimates |
|---|---|---|
| On-ramp fees | Fees for converting fiat money to payment stablecoins by digital assets services providers including wallets, exchanges, among others which conduct user verifications and | 0.00% - 1.00% |

---

[21] Uniswap Labs and Circle Internet Financial, On-chain FX and cross-border payments, 2023.

| | link to fiat payment systems. This process relies on local fiat payment rails such as ACH or SEPA. | |
|---|---|---|
| Exchange fees | Fees paid to liquidity providers for converting stablecoins. | 0.01% - 0.05%. |
| Network transactions fees | Fees paid to network validators associated with computation costs in automated market makers, exchanging payment stablecoins, and sending from wallet to wallet. | $0.35 - $5.00 |
| Off-ramp fees | Fees from converting payment stablecoins for fiat money for the recipient. If the recipient continues to use payment stablecoins domestically, this can be avoided. | 0.00% - 1.00% |
| **Total cost of sending remittances using stablecoins** | | **From 0.01% + $0.35 to 2.05% + $5.00** |

Source: On-chain Foreign Exchange and Cross-border Payments, 2023

**Stablecoins are a type of digital currency currently used for cross-border payments; a Central Bank Digital Currency (CBDC) could also play a key role in cross-border payments in the future.** A stablecoin is a category of digital assets. Its main feature is that it is fully backed by a fiat currency and/or low-risk and highly liquid assets such as short-term government bonds. Design considerations are crucial to ensuring that a stablecoin has adequate reserves, appropriate custody procedures, redemption rights, disclosures and transparency requirements, among others. However, central banks in LAC highlighted that the growth of USD stablecoins risks informal dollarization if emerging economies rely primarily on USD-pegged stablecoins. Stablecoins are the main type of digital currency used for cross-border payments; however, CBDCs may serve the same purpose in the near to medium term should CBDC implementation and adoption accelerate.

**Box 3. Easing cross-border payments**

On- and off-ramp fees

In Spring 2023, financial entities **FinClusive** and **Anclap** announced the results of a limited-scope pilot focused on the US-Colombia remittance corridor, utilizing blockchain technology.

For this particular corridor, the average cost of sending remittances via traditional methods is almost 6 percent (5.98 percent) of the amount sent, according to [data](#) from The World Bank. Based on the results of this pilot, the average cost for remittances sent via blockchain, utilizing stablecoins, is roughly half that figure (1.60 - 2.50 percent, depending on liquidity) for this specific payment corridor and could be further reduced as transaction volume increases.

In addition, the pilot highlights the importance of interoperability, showing that blockchain-based solutions can integrate seamlessly with local banking and payments infrastructure, ensuring a convenient and user-friendly experience. While these results are based on a small sample size, it reflects the time and cost for blockchain-based real world payments between individuals.

**A cross-border CBDC could improve the cost and use of remittances.** According to the 2022 CEMLA survey data, one of the main perceived benefits of a CBDC is greater speed and cost reduction for payments and transfers, including remittances.[22] In addition, some central banks mentioned that the local payments system with neighboring countries can also benefit from the implementation of a CBDC. Other central banks also highlighted that a CBDC system could improve financial inclusion, favor interoperability, improve agility in electronic payments, and provide relevant information on consumption and market trends, for example, in segments like agriculture and international remittances.

*Assessment and comparison to other payment solutions*

| Element | Cards scheme | Payments originated through mobile devices | Assessment: Digital Currency system |
|---|---|---|---|
| a) Cost and affordability | Domestic (for merchants): **2** percent for debit cards Up to **6** percent for credit cards[23] | Domestic (for merchants): **0.81** percent to **1.3** percent[24] | LOW IMPACT: Although there are currently some CBDC implementations globally, there is no clear evidence that suggests that digital currencies are reducing the fees for accepting digital payments by retailers. |
|  | International: **5.07** percent to fund a remittances with a card | International: **4.36** percent to fund a remittance with mobile money | HIGH IMPACT: **2** percent. |

b) **Availability**

**In addition to lower costs, digital currencies for payments provide real-time settlement, availability, transparency, and further liquidity.** Access to an individual's digital currency is not restricted by market or banking hours and payment transactions are generally settled in real-time. Transactional data on blockchains are also traceable, but not linked to personal identifiable information. Further, the digital currency liquidity market has been steadily growing,[25] filling out gaps when funds are scarce.

*Assessment and comparison to other payment solutions*

| Element | Cards scheme | Payments originated through mobile devices | Assessment: Digital Currency system |
|---|---|---|---|
| b) Availability | Settlement of card payments are available **only on business days**. | Settlement of mobile money payments are available **only on** | HIGH IMPACT: Digital currency systems are currently available 24/7. |

---

[22] Gerardo Hernandez-del-Valle, Alicia Chua, Bernardo Panigua, CEMLA's survey on Central Bank Digital Currencies in Latin American and the Caribbean, 2023.
[23] For Brazil merchants fees are lower.
[24] For Argentina where data is available.
[25] For example, in 2022, financial entity Arf introduced a short-term liquidity solution using USDC for cross-border payments. Within eight months of its launch, the company has extended over $270 million in loans to financial institutions, resulting in a cumulative on-chain USDC volume of $370 million.

| | This impacts merchants that do not receive funds immediately during the weekend. | **business days**, with the exception of P2P transfers. This impacts merchants that do not receive funds immediately during the weekend. | |

### c) Accessibility and convenience

**A robust digital currency system may contribute to increasing the distribution and access of money in remote and rural areas; however, connectivity is a challenge in LAC.** Although digitalization has improved the possibility to access the financial infrastructure, the distribution of physical money is still a challenge in some countries of LAC. Physical money is still used in cash-based economies, especially for those individuals and businesses that operate in the informal sector. A robust digital currency system could help distribute and expand the access of money in remote and rural areas. Connectivity is a precondition for the development of a digital currencies system. Central banks of LAC highlighted that a CBDC system can contribute to the distribution of central bank money.

*<u>Assessment and comparison to other payment solutions</u>*

| Element | Cards scheme | Payments originated through mobile devices | Assessment: Digital Currency system |
|---|---|---|---|
| c) Accessibility and convenience | Although adoption by merchants of cards is increasing, generally merchants still do not accept card payments in remote and rural areas due to the high cost, among others.<br><br>Payment services providers developed payment solutions using technology like Near Field Communication (NFC), which allows consumers and merchants to use and accept digital payments without internet or telecommunication connectivity by relying on a deferred settlement system that completes a transaction when network access becomes available. | Although adoption by merchants of mobile money is increasing significantly in some jurisdictions, generally merchants still do not accept mobile money payments in remote and rural areas. | MEDIUM IMPACT: Although the private sector in the digital currencies space is developing offline solutions, products are still in an early stage of development and are not designed to function 100% offline yet. Developing the digital infrastructure in LAC is a precondition for accessibility. |

### d) Scalability

**A digital currency system can adopt a wide variety of design characteristics depending on the public policy objectives and system performance demands.** It is important that policy and technical design and implementation are not conducted in isolation from each other. Depending on the local context and volume of transactions in a national payment system, demand for the system capability can vary. For example, the performance demands required for a country like Brazil differ from a smaller country in the region. A robust digital currency system should process the volume of transactions per second required by the performance demands defined by each national payment system operator in LAC.

*Assessment and comparison to other payment solutions*

| Element | Cards scheme | Payments originated through mobile devices | Assessment: Digital Currency system |
|---|---|---|---|
| d) Scalability | The card scheme can process 65,000 transactions per second.[26] | Mobile money solutions can process 544,000 transactions per second.[27] | HIGH IMPACT: **1.7 million** transactions per second with 99% of transactions durably completing in under a second and the majority of transactions completing in under half a second, according to preliminary CBDC experimentation. |

### e) Programmability

**A well-designed digital currency system should possess the flexibility to tailor the distribution of digital currencies to incorporate specific conditions required by the local context.** Under specific circumstances, it may be necessary to program payments to achieve a policy objective. For example, programmable payments can be relevant for the disbursement of social assistance and emergency relief programs. Programmable payments can also be utilized for automated initiation of payments on confirmed receipt of goods or routing payments directly to the local tax authorities. A robust digital currency system should be flexible in order to incorporate any specific requirements and/or conditions.

*Assessment and comparison to other payment solutions*

| Element | Cards scheme | Payments originated through mobile devices | Assessment: Digital Currency system |
|---|---|---|---|
| e) Programmability | Although card schemes can be tailored to incorporate transaction limits and similar features, they do not have extensive programmability functionalities. | Although payments originated through mobile money can be tailored to incorporate transaction limits and similar features, they do not have extensive programmability functionalities. | HIGH IMPACT: Current digital currency systems and preliminary CBDC experimentation showcase that conditional payments through programmability are feasible. |

---

[26] Visa, Fact sheet
[27] Atlantic Council, A Report on China's Central Bank Digital Currency: the e-CNY

# 5. High-level recommendations for leveraging digital currencies for financial inclusion in LAC

Based on the available quantitative and qualitative data and the assessment of the current digital currencies system using the proposed framework, this section presents high-level recommendations that central banks and financial service providers can take to design, test and implement a robust system that can leverage digital currencies for financial inclusion.

| High-level recommendations for the public sector | |
|---|---|
| High-level recommendation 1 | Policy goal: Increase access to an account |
| High-level recommendation 2 | Policy goal: Reduce merchants' acceptance costs |
| High-level recommendation 3 | Policy goal: Develop instant payment settlement |
| High-level recommendation 4 | Policy goal: Reduce the cost of cross-border payments |
| High-level recommendation 5 | Policy goal: Reduce the use of cash and bring individuals and MSMEs into the formal economy |
| High-level recommendation 6 | Policy goal: Expand the availability and access to central bank money |
| High-level recommendation 7 | Policy goal: Develop a legal and regulatory framework for digital currencies |
| High-level recommendation 8 | Policy goal: Disburse programmable payments |
| High-level recommendation 9 | Policy goal: Foster international cooperation and harmonization |
| Private sector considerations: Design and product development recommendations | |
| High-level recommendation 10 | Apply a human-centered approach to improve the customer journey for vulnerable groups |
| High-level recommendation 11 | Consider adoption as a design element |
| High-level recommendation 12 | Develop offline solutions |
| High-level recommendation 13 | Utilize a phased implementation approach to deploy digital currencies solutions |

### a) High-level recommendations for the public sector: Policy goals to be considered when designing a digital currency system

This subsection aims to assist central banks and policymakers in defining policy goals and motivations for designing a digital currency or CBDC system.

*High-level recommendation 1: Policy goal: Increase access to an account*

Data shows that in some countries in LAC the rate of account ownership is significantly lower than in other regions globally. According to the World Bank Global Findex, the account ownership rate in LAC amounts to 72.9 percent. However, the account ownership for emerging economies in other regions is higher. For instance, the rate is 80.8 percent in East Asia and the Pacific. In addition, data shows that the account ownership in some LAC countries is lower than 30 percent.

A well-functioning payment instrument is an important precursor that incentivizes uptake of accounts. The ability to send and receive digital payments via a robust digital currency or CBDC system can broaden the use of accounts to store funds and make transactions. Mobile money solutions such as digital wallets, P2P transactions through mobile devices, and QR code payments have proven to be effective tools that have resulted in the increased use of digital payments. As a next step, banks and non-banks should be prepared to offer accounts that serve mobile money users. Regulators could also assess implementing a risk-based approach in terms of anti-money laundering regulations that would simplify the onboarding process of new customers. The World Bank Global Findex shows that from 2017 to 2021 mobile money accounts increased from 5.2 percent to 23.4 percent. A well-developed digital currency system could trigger further access, specifically for vulnerable groups and small businesses operating in the informal economy in LAC.

*High-level recommendation 2: Policy goal: Reduce merchants' acceptance costs*

According to central bank data from the LAC region, the cost for merchants to accept debit and credit cards varies between 0.81 and 6 percent, depending on the jurisdiction. In countries like Argentina, merchants that accept mobile device payments via QR codes and mobile device PoS terminals are subject to lower fees – between 0.8 and 1.3 percent – illustrating that mobile money solutions are reducing costs for retailers. However, these solutions typically require the intermediation of several financial entities: QR code service providers, aggregators, digital wallet service providers, card issuers, acquirers (both for cards and mPoS), and payment processors.

A well-designed digital currency system can further reduce intermediation and the cost of processing payments. Policymakers should encourage greater uptake of mobile payment solutions in the near-term while also designing a framework for digital currencies that can eliminate the need for financial intermediaries to make payments and transactions more affordable. Furthermore, central banks should assess regulatory measures that reduce the interchange fee charged by card-issuing entities.

*High-level recommendation 3: Policy goal: Develop instant payment settlement*

Based on responses from central banks in LAC countries, central banks consider their real time gross settlement (RTGS) systems to be efficient, but believe features like 24/7 functionality through a CBDC system and instant settlement can improve them.

Currently, national payment systems do not offer 24/7 days availability. Generally, settlement occurs only during business days. This brings liquidity risks and credit exposure to the financial system but, most importantly, it impacts users and mainly merchants which are unable to access funds instantly.

A robust and efficient digital currency system can contribute to settling payments real-time and reducing liquidity and credit exposure risks. This would also help reduce the cost of merchant acceptance and drive digital payments adoption.

*High-level recommendation 4: Policy goal: Reduce the cost of cross-border payments*

The global average cost of sending USD 200 is 6.25 percent. However, the average cost of remitting from G20 countries is higher, at 6.47 percent. Data shows that the most affordable method for sending a remittance is through a mobile money platform; however, it is still higher than the 3 percent SDG target, at 4.36 percent.

Digital currencies contribute to a reduction in the cost of cross-border payments. The total cost of remittances using stablecoins is estimated at 2 percent. Therefore, the design considerations of a stablecoin are crucial to ensure that a stablecoin can be used for payments, both cross-border and retail. Policymakers and regulators should monitor the development of cross-border payments solutions and incentivize adoption of digital currencies for payments through regulatory clarity.

*High-level recommendation 5: Policy goal: Reduce the use of cash and bring individuals and MSMEs into the formal economy*

LAC economies are still largely cash-based societies, unlike developed economies that have evolved into the digital space. Many actors use cash in the informal economy to avoid traceability and tax obligations.

Digital currencies can support MSMEs and individuals transition from the informal to the formal economy. Central banks from the LAC region highlighted that a retail CBDC could reduce the use of cash and develop a more efficient payments system.

*High-level recommendation 6: Policy goal: Expand the availability and access to central bank money*

Some Caribbean countries are composed of small, interconnected islands. Due to these conditions, central banks in the region struggle to distribute cash to individuals and MSMEs. Traditional downstream facilitators like bank branches and ATMs are often unreliable because of natural disasters like hurricanes that impact physical and network connectivity.

A well-designed and robust digital currency system could improve the distribution and access of central bank money through digital wallets and innovative solutions that do not require bank branches or agents. A digital wallet solution also improves the distribution of funds for social assistance programs and relief schemes. Connectivity and access to digital infrastructure should be a precondition for the development of a reliable financial system.

*High-level recommendation 7: Policy goal: Develop a legal and regulatory framework for digital currencies*

While innovative digital solutions can address inclusion barriers, new business models – especially financial entities offering and intermediating a digital currency – can pose risks to consumers without proper oversight and supervision. There is even a question as to whether financial entities facilitating digital currencies transactions are authorized to engage in this activity under the current legal and regulatory framework in many LAC countries. Regulatory sandboxes could be a tool that policymakers and regulators could use to address these risks and better understand how these new business models work.

For example, a well-designed legal and regulatory framework for stablecoins should require adequate reserves backed by fiat money or a combination of fiat money and high-quality liquid assets such as short-term government bonds, held in the same currency as the token, which are in custody in regulated financial entities, and subject to independent third-party attestation that is made publicly available on a regular basis. Stablecoins arrangements should include disclosure and transparency requirements to ensure consumer protection and privacy and provide a clear guarantee of redeemability. This would drive the development of a robust financial sector to distribute digital currencies.

There is also a question as to whether central banks have the authority to issue a CBDC under the current legal framework in each jurisdiction. Although several central banks already have a broad mandate to issue the local currency, central banks may need to obtain a new mandate by legislative bodies such as congresses or other governmental authority to issue a CBDC.

*High-level recommendation 8: Policy goal: Disburse programmable payments*

A digital currency system can support the following programmable payments use cases: (a) automated initiation of payments on confirmed receipt of goods, (b) routing tax payments directly to the tax authorities at point of sale; (c) interest payments on securities, and (d) social payments for relief programs or emergency cash transfers.

According to the responses from the CEMLA survey, central banks in LAC consider programmability as a fundamental functionality for a potential CBDC system. Programmable payments could achieve greater efficiency, transparency and resiliency, specifically for the disbursement of relief programs, emergency assistance, and social payments.

*High-level recommendation 9: Policy goal: Foster international cooperation and harmonization*

Central banks and policymakers in LAC should work together to share knowledge and foster cooperation of digital currency regulatory bodies and harmonization of standards at the international level. Coordination among relevant authorities at the domestic level is also necessary to drive this process and ensure standards are developed in a consistent manner.

According to the responses from the CEMLA survey, although each country has its own characteristics and context and a unique CBDC system would be required, there are several common topics of interest, such as financial inclusion and remittances, that can benefit from greater coordination and cooperation. This could be channeled through standard-setting bodies such as the FSB, BIS CPMI, IOSCO and FATF that are currently working to achieve harmonization of regulatory and technical requirements.

### b) Private sector considerations: Design and product development recommendations

This subsection aims to inform product development for digital currencies for the private sector.

*High-level recommendation 10: Apply a human-centered approach to improve the customer journey for vulnerable groups*

Financial entities should apply human-centric design techniques when designing digital currencies solutions to simplify the user experience. Incorporating data analytics and customer insights in the product development process can ease chokepoints like customer onboarding and improve the overall consumer experience. More importantly, a human-centered approach can tackle the unique needs of vulnerable groups, especially the unbanked and underbanked.

Financial entities should design tailored solutions for specific target groups, taking into consideration specific needs and circumstances, instead of deploying a one-size-fits-all solution. Financial entities should also consider and reconcile conflicting interests of different users. For instance, it is probably easier for a consumer to make a payment with a credit card rather than scanning a QR code since the latter involves further steps than making a contactless payment with a credit card. However, it could be more beneficial for a merchant to accept a QR payment rather than a credit card payment due to the costs associated with each transaction. An easy-to-use digital currency solution would only require a few steps or "clicks" to proceed with a transaction.

*High-level recommendation 11: Consider adoption as a design element*

A robust digital currency system requires adoption by both consumers and merchants. Adoption is a key challenge facing systems in an early stage of development. This applies to CBDCs and privately-issued digital currencies.

Adoption should be incorporated in the design and product development process. Financial entities should understand consumer frictions and articulate the problem that a newly-launched product is resolving. Furthermore, financial entities should create the appropriate incentives and environment for consumers to transition from traditional to innovative financial products by designing products to be easy-to-use and more affordable. For example, if a digital currency wallet is customer-friendly for the consumer and cheaper for the retailer to accept, adoption would be a natural and seamless process that would create customer loyalty and deepen the financial sector.

*High-level recommendation 12: Develop offline solutions*

Consistent connectivity and access to network infrastructure is one of the main challenges in the development of a digital currency system. According to the CEMLA survey responses, lack of connectivity is a common issue in rural or remote areas in LAC. However, in several countries in the region, the rate of smartphone adoption is increasing. For example, mobile phone penetration in Argentina reached 88.1 percent in 2022.[28]

Developers leverage the growth of mobile phone adoption by creating payment solutions like NFC, which allows consumers and merchants to use and accept digital payments without internet or telecommunication connectivity. Offline payments allow users to transact without connectivity by using a deferred settlement system that finalizes when network access becomes available. There are different models for value exchange and settlement of offline transactions. Value exchange and settlement between two parties can happen entirely offline, or users can synchronize with the ledger prior to an exchange. Offline payments functionality is crucial in economies that may commonly face operational failures and disruptions like electrical outages. Given the rise in adverse impacts from climate change, offline functionality serves a key role in allowing users to make payments 24/7 even in the face of destructive natural disasters.

Offline payment solutions are in an early development stage and still face challenges like double spending, which occurs when users spend the same offline value more than once. Financial entities should continue developing services that provide instant offline settlement and address settlement risks like double spending.

*High-level recommendation 13: Utilize a phased implementation approach to deploy digital currencies solutions*

Businesses should apply a phased approach for the design and launch of digital currency solutions. First, firms should collect and analyze existing data to identify a specific problem.

---

[28] Banco Central de la República Argentina, *Informe de Inclusión Financiera*, 2023.

Financial entities should then engage potential customers to obtain further insights to better understand the target market.

Next, firms should use these findings to inform product development. Firms should prototype and pilot the proposed solution in a controlled environment with a small group (usually called "friends and family"). Firms should gradually expand the user base to obtain more data on the efficacy of the product. This step would also involve fixing any issues identified during testing. This could be done in collaboration with regulators under regulatory sandboxes.

The implementation phase can begin if testing results are positive. This involves the launch of the digital currency solution in the market, which is usually complemented with specific marketing campaigns to drive adoption and usage by the target group.

## 6. Conclusion

Despite increased uptake in global account ownership, LAC countries continue to struggle providing adequate access to accounts and the formal financial system. Between 2017 and 2021, usage of digital payments increased significantly in LAC countries, from 45.1 percent to 65.1 percent. However, although some countries in the region introduced incentives to increase the acceptance by merchants of digital payments, fees for debit and credit cards remain high.

Cross-border payments face similar challenges. The cost of sending funds across borders is still high with a global average cost of 6.25 percent for every USD 200. Technology contributes to a reduction in the average cost of remittances. The cheapest method for funding a remittance transaction is mobile money; however, the cost is still higher than the 3 percent SDG target.

A CBDC could reduce the use of cash and make payments more affordable for consumers and merchants. CBDCs may contribute to increasing the distribution and access of central bank money in remote areas and reducing the cost of printing central bank money. However, connectivity and access to digital infrastructure is an obstacle that needs to be addressed for the development of a robust digital currency system.

This report suggests a framework for assessing the contribution of digital currencies and design considerations to promote financial inclusion in LAC. The framework is comprised by the following elements:

   a) Cost and affordability
   b) Availability and settlement time
   c) Accessibility and convenience
   d) Scalability
   e) Programmability

Although digital currencies development is at an early stage, a well-designed system could reduce the cost of domestic and cross-border payments, improve the settlement of transactions to achieve real-time payments, expand the accessibility of central bank money, incorporate programmable payments and achieve system performance demands. The public and private sector should work together to ensure that the design, development and implementation of a digital currency system addresses policy goals and assists creating financial products that are easy-to-use to drive adoption.

## Annex 1. Questions from survey

1. Country generalities.
    a. Geography
    b. Political
    c. Economic
    d. Technological
    e. Sovereign currency

2. What other factors do you think will benefit or affect the implementation of a CBDC in your country (e.g. the banking sector will see a CBDC as an unfair competitor, lack of financial literacy could affect the CBDC penetration, etc.)?

3. What are the Central Bank's objectives with respect to the banking and financial system?

4. Which other authority shares the responsibility of the banking and financial system (e.g., financial market authority or the Finance Ministry)?

5. Is the Central Bank autonomous?
    a. How much autonomy does the central bank have in the design, development and implementation of a CBDC?
    b. Is there any current barrier impeding the Central Bank to issue a CBDC? Has this been discussed internally?
    c. Has your central bank explored the possibility of designing and implementing a CBDC in some form in the past?

6. What is the current design of your monetary system and what are the problems with it (e.g. the Central Bank relies on the private sector for deposit and lending activities by regulation)?

7. Where does the central bank sit with respect to issuance of sovereign currency and the country's government (e.g. the Central Bank is the sole issuer of the sovereign currency)?

8. How does the central bank interface with banks and other regulated financial entities (e.g. the Central Bank regulates and oversees commercial banks for their financial intermediation activities, the Central Bank gives them access to the payment systems)?

9. What is the state of the technology and infrastructure for payments between the Central Bank and banks and other regulated financial entities (e.g. all retail payment systems are run by the private sector, the Central Bank only runs the RTGS)?
    a. What are the main weaknesses that the Central Bank can identify from these arrangements?
    b. Does the central bank interface directly with payments service providers (e.g. e-money companies)?

10. How is your wholesale payment system, do you consider it developed or does it need to be made efficient (e.g. there is enough interbank market activity and intraday liquidity in the system is managed smoothly by banks and other regulated financial entities with access to central bank money)?

11. How much competition is there in your financial and payment services industry?

    a. Are there issues with dominance of certain systems or providers that are causing a lack of innovation (e.g. a private ACH forbidding nonbanks to become participants, banks set fees for transactional services like deposits and payment accounts)?

    b. How can you assess the operation and coverage of the domestic retail payment system? Do you consider it developed or does it need to be made efficient (e.g. in terms of speed, costs, access by nonbanks, etc.)?

12. Do the current cross-border payment services meet all needs by their different users (e.g. remittances recipients have immediate access to their funds, the total fees are below 10% of the total amount sent)? What other services can you satisfy with cross border transactions?

13. What are the major issues for the monetary policy and financial stability that can be jeopardized by the current state of the domestic and cross-border payments infrastructures?

    a. Are the payment channels/instruments (e.g. RTGS services) for banks easily available and developed to contribute to an effective monetary policy transmission (e.g. central bank operations can be settled seamlessly in the payment infrastructure?

    b. How is financial stability and integrity ensured with the current infrastructure (e.g. in case of liquidity shortage by banks, intraday and overnight liquidity is available at the critical payment systems)?

14. Regarding capabilities and legal requirements, what are the technical competencies and experience of running payment infrastructure? Is the same for the Central Bank and the private sector?

    a. Is there any relevant experience of working with emerging technologies (DLT)?

    b. How would new forms of money like e-money systems fit (or not) under current legal designations?

15. How are the following forms of money (or emerging forms of money) utilized in your economy and what are the trends you are observing?

    a. Cash (e.g. it is used for small transactions, is the form of money most used in all type of transactions)

    b. Bank deposits (e.g. it has low levels of access, it is only used for saving purposes, it competes with cash)

    c. Central bank reserves (e.g. it is only accessible for banks and other regulated financial entities)

    d. E-money (digital payments) (e.g. it is relatively new and not fully adopted, it competes with bank deposits, it is under implementation)

    e. Cryptocurrencies (e.g. they are not authorized as a form of money, they are not present in the country, they are commonly used as a financial asset for investment purposes)

16. What is the prevalence of cash in your economy (e.g. versus other forms of money like

e-money, debit cards, etc.)?

    a. How expensive is to produce and manage cash?

    b. How does this relate to objectives on financial inclusion or efficiency in the financial and payments services industry?

17. What is the current legislation around financial consumer protection (e.g. in terms of privacy and identity, resolution of disputes, etc.)?

18. Regarding the main component, what would be the consensus for your Central Bank?

    a. Network Topology Consensus

    b. Immutability and Failure Tolerance Consensus

    c. Gossiping consensus (Global-local)

    d. Consensus by agreement

19. What are the transition capabilities that your central bank could have or have?

    a. Data structure in the Blockheader

    b. Transaction model

    c. Server storage

    d. Block storage

    e. Limits to scalability (number of transactions, number of nodes, number of users and block confirmation time)

20. What would be the configuration of your native currency/Tokenization?

    a. Native asset

    b. Asset supply management

    c. Tokenization

21. What would be the extensibility you are looking for?

    a. Interoperability

    b. Intraoperability

    c. Governance

    d. Script language

22. What would be essential for your bank in terms of security and privacy? Do you prefer:

    a. Data encryption (SHA-2, SK-SNARKS or other, if so which one)

    b. Data privacy (Built in data privacy, add-on data privacy or other, if so which one)

23. In terms of identity management, what do you prefer?

    a. Access layer and control layer

    b. Identity layer

24. For the encoding and charging and rewarding system, which is most desirable for your system?

a. Codebase

   b. Charging and rewarding system

25. Regarding your country's infrastructure and economy?

   a. How good is infrastructure for:

   b. Are existing legal and regulatory requirements compatible with the issuance of CBDC (e.g. the Telecommunications Authority has a good degree of coordination with the Central Bank and the financial industry to set standards)? What potential legal or regulatory obstacles exist as limitations to implementing a CBDC?

26. Is there demand and interest in CBDC between the domestic financial and payments industry? Have you surveyed this? Relatedly, which parts of the public or private sector require it, in your economy? What is the support that the Central Bank senior management gives to this issue?

27. In relation to question 19, what are the risks or fears of implementing a CBDC that you foresee in your country? Are they well-studied risks or do you consider that they could deserve special attention? How is the resilience of your payment system?

28. Thinking about all aspects of this exploratory analysis. What are areas and points you think/expect that CBDC can be useful for your country?

29. What is the knowledge, experience or do you have experts studying CBDC? Is there a CBDC related research agenda in your country? How do you think we could support you for the analysis of a CBDC? What do you expect from us in this exploratory research project?